\documentclass[authoryear,final,12pt]{elsarticle}

\usepackage{amssymb}

\usepackage{times}
\usepackage{blindtext, graphicx}
\usepackage{colortbl}
\usepackage{tikz}
\usepackage{microtype}
\usepackage[short,nodayofweek,level,12hr]{datetime}

\usepackage{booktabs}
\newcommand{\bi}{\begin{itemize}}
\newcommand{\ei}{\end{itemize}}
\usepackage{slashbox}
\usepackage{array}
\newcolumntype{P}[1]{>{\centering\arraybackslash}p{#1}}

\usepackage{balance}
\definecolor{Gray}{rgb}{0.88,1,1}
\definecolor{Gray}{gray}{0.85}
\definecolor{lightgray}{gray}{0.8}
\usepackage{multirow}

\usepackage{subfig} 

\usepackage{makecell}
\usepackage[linesnumbered,ruled,vlined]{algorithm2e}
\SetKwProg{Fn}{Function}{}{}

\usepackage{amsmath}

\usepackage{wrapfig}

\usepackage{enumitem}
\setlist[itemize]{leftmargin=*}
\setlist[enumerate]{leftmargin=*}
\setlist[enumerate]{nosep}

\usepackage[framed]{ntheorem}
\usepackage{framed}
\usepackage{tikz}
\usetikzlibrary{shadows}
\theoremclass{Lesson}
\theoremstyle{break}

\tikzstyle{thmbox} = [rectangle, rounded corners, draw=black,
fill=Gray!20, drop shadow={fill=black, opacity=1}]

\newshadedtheorem{lesson}{Finding}

\usepackage[tikz]{bclogo}
\newenvironment{RQ}[1]%
{\noindent\begin{minipage}[c]{\linewidth}%
\begin{bclogo}[couleur=gray!20,%
 arrondi=0.1,%
 logo=\bctrombone,%
 ombre=true]{~#1}}%
{\end{bclogo}\end{minipage}\vspace{2mm}}

\def\checkmark{\tikz\fill[scale=0.4](0,.35) -- (.25,0) -- (1,.7) -- (.25,.15) -- cycle;} 

\newcommand{\IT}{FAST$^2$}

\usepackage{xcolor}
\usepackage{natbib}
\usepackage[colorlinks=true]{hyperref} 

\journal{Expert Systems with Applications}

\begin{document}

\begin{frontmatter}




\title{{Assessing Expert System-Assisted Literature Reviews With a Case Study}}


\author[rit]{Zhe Yu}
\ead{zxyvse@rit.edu}
\author[UA]{Jeffrey C. Carver} 
\ead{carver@cs.ua.edu}
\author[ncsu]{Gregg Rothermel} 
\ead{gregg.e.rothermel@gmail.com}
\author[ncsu]{Tim Menzies}
\ead{timm@ieee.org}
\address[rit]{Rochester Institute of Technology, Rochester, NY, USA.}
\address[ncsu]{North Carolina State University, Raleigh, NC, USA.}
\address[UA]{University of Alabama,  Tuscaloosa, AL, USA.}

\begin{abstract}Given the large numbers of publications in software engineering, frequent literature reviews are required to keep current on work in specific areas. One tedious work in literature reviews is to find relevant studies amongst thousands of non-relevant search results. In theory, expert systems can assist in finding relevant work but those systems have primarily been tested in simulations rather than in application to actual literature reviews. Hence, few researchers have faith in such expert systems. Accordingly, using a realistic case study, this paper assesses how well our state-of-the-art expert system can help with literature reviews. 

{The assessed literature review aimed at identifying test case prioritization techniques for automated UI testing, specifically from 8,349 papers on IEEE Xplore.} This corpus was studied with an expert system that incorporates an incrementally updated human-in-the-loop active learning tool. Using that expert system, in three hours, we found 242 relevant papers from which we identified 12 techniques representing the state-of-the-art in test case prioritization when source code information is not available. These results were then validated by six other graduate students manually exploring the same corpus. Without the expert system, this task would have required 53 hours and would have found 27 additional papers. That is, our expert system achieved 90\% recall with 6\% of the human effort cost when compared to a conventional manual method. Significantly, the same 12 state-of-the-art test case prioritization techniques were identified by both the expert system and the manual method. That is, the 27 papers missed by the expert system would not have changed the conclusion of the literature review. 

Hence, if this result generalizes, it endorses the use of our expert system to assist in literature reviews.
\end{abstract}



\end{frontmatter}

\section{Introduction}
\label{sect: Introduction}

New papers are being published every day, and in increasing numbers. Knowing what other researchers have done to address a problem has become equally important as, if not more important than, providing a novel solution. However, it is also increasingly difficult to stay current with what other researchers are doing. For example, when searching for work on test case prioritization (TCP) on IEEE Xplore\footnote{\url{https://ieeexplore.ieee.org}}, 2,704 results would have been returned in 2009, while in 2019, that number has grown to 8,349. As a result, finding an efficient way to conduct literature reviews and extract useful information from thousands of papers has become a crucial problem for researchers.

To address this problem, software engineering researchers have introduced Systematic Literature Reviews (SLRs) \citep{kitchenham2004evidence,1377125}. Following a set of guidelines, researchers conducting SLRs manually examine all of the papers relevant to a set of research questions and summarize the research area. Other researchers can then obtain a general idea about current activity in their field of interest by reading the published SLRs. However, SLRs are conducted infrequently because of their labor-intensive and time-consuming nature~\citep{Yu2018}. As a result, when researchers explore a specific problem, they often find that existing SLRs are outdated and they need to carry out their own literature reviews.

{In theory, expert systems can assist humans in SLRs and reduce the human effort required. However, those expert systems have primarily been tested in simulations only, rather than in application to actual literature reviews. Consequently, few researchers have faith in such expert systems. To address the above problem, this paper assesses whether it is useful to apply expert systems to actual literature reviews with a literature review case study.}

\begin{wrapfigure}{r}{2.4in}
\includegraphics[width=2.35in]{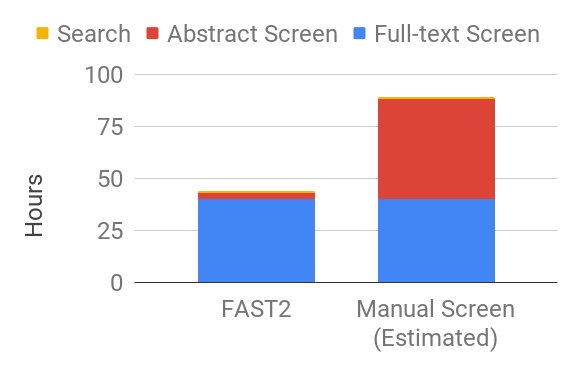}
\caption{Human hours required for conducting the literature review. The targets for the abstract screen are to identify 90\% of relevant papers.}\label{fig:compare}
\end{wrapfigure}

We (the authors of this article) were faced with the same literature review problem when exploring the state of the arts in applying TCP techniques to automated UI tests~\citep{yu2019terminator}. Given the specific nature of automated UI tests, only test history, test description, and test results information are available for prioritizing the test cases. Therefore, we needed to conduct our own literature review searching for TCP techniques that rely only on the available information. One tedious task we faced in doing this was to find all the relevant TCP papers from among 8,349 search results on IEEE Xplore, using their titles and abstracts. In our estimation, this would require around 53 human hours, as shown in Figure~\ref{fig:compare}. To reduce this time, we applied a state-of-the-art human-in-the-loop expert system called {\IT}~\citep{Yu2018,Yu2019} to assist in selecting relevant papers. By screening and labeling 470 papers suggested by the expert system in three hours, the first author identified 242 relevant papers for full-text reviewing. The algorithm indicated that those 242 relevant papers constituted 91\% of all the relevant papers in the
8,349 search results. Finding more, however, would require much more human effort. This was an impressive result because 50 human hours, which is 94\% of the original time required, can be saved by sacrificing 9\% recall (missing 24 relevant papers). Considering the 40 hours required for full-text screening and the 1 hour required for search, using {\IT} saved about half of the time and effort required for the literature review, as shown in Figure~\ref{fig:compare}.

Since this was the first time {\IT} was applied in a real literature review, we wished to validate its result. Enlisting six other graduate students to manually screen a subset of the candidate papers and perform full-text reviewing of the missing relevant papers from {\IT}, we explored the following research questions.

\textbf{RQ1: What percentage of relevant papers does {\IT} actually retrieve?}
The {\IT} selection process included 85\% of the relevant papers with human errors contributing to 5\% (12) of the missing papers. That said, {\IT} was in fact responsible for missing 10\% (27) of the relevant papers, which is very close to its estimation of having missed 9\% (24) of the relevant papers. 

\textbf{RQ2: What information is missing in the final report because of {\IT}?}
The distributions of the missing relevant papers are different from the distribution of papers that {\IT} included. This suggests that {\IT} introduced a sampling bias to the relevant papers it identified. On the other hand, with the missing relevant papers added into the included papers, the distribution of the included papers remained roughly unchanged. Additionally, the overall conclusions of the literature review (12 TCP techniques identified suitable for automated UI testing) were not affected by the missing relevant papers (39). 

In conclusion, {\IT} was able to include 90\% of the relevant papers, as it claimed, and the 242 papers it included were sufficient for our literature review. The set of TCP techniques identified through the literature review was used as a baseline in our published work on prioritizing automated UI tests~\citep{yu2019terminator}. We believe that saving 50 hours in selecting relevant papers with 10\% of the relevant papers omitted is a worthwhile result. This may also extend to other situations in which SLRs are conducted. When conducting studies such as system mappings, however, where the actual number and distribution of papers matters, Active learning-based expert systems such as {\IT} may need to be avoided until the sampling bias issue has been resolved.

The main contributions of this paper are as follows:
\bi
\item
We conducted the first systematic literature review using {\IT}, a state-of-the-art expert system for assisting with relevant paper selection. This is also among the very few SLRs assisted by machine learning algorithms or expert systems.
\item
We validated the use of {\IT} in an SLR and observed its strengths and weaknesses {via a controlled experiment of only humans performing the same relevant paper selection task.}
\item
We scripted our entire SLR process and provided it in a public Github repository\footnote{\url{https://github.com/fastread/SLR_on_TCP}} so that other researchers can use the data for reproduction and improvements on machine-learning-assisted primary study selection.
\ei

The rest of this paper is structured as follows.
Section~\ref{sect: Background} presents background material and related work. Section~\ref{sect: Research method} reports every detail of the performed literature review case study. Section~\ref{sect: validate} reports the validation results of {\IT} and answers the research questions via a controlled experiment. Section~\ref{sect: Conclusions and Future Work} concludes the paper and discusses potential future work.


\section{Background and Related Work}
\label{sect: Background}

In this section, we provide background information on systematic literature reviews, and
the state-of-the-art machine learning algorithms and expert systems supporting primary study selection.

\subsection{Systematic Literature Reviews}

Systematic Literature Reviews (SLRs) have become a well established and widely
applied review method in Software Engineering since Kitchenham, Dyb{\aa}, and
J{\o}rgensen first adopted them to support evidence-based software engineering in
2004 and 2005~\citep{kitchenham2004evidence,1377125}. SLRs employ a defined search strategy, and an
inclusion/exclusion criterion for identifying the maximum possible relevant literature. As a result, compared to traditional literature reviews, an SLR provides thorough, unbiased and valuable summaries of the existing information on specific research questions. However, SLRs also require much more human effort than traditional reviews (weeks to months of work as reported in~\citep{Yu2018}). Therefore, SLRs cannot be conducted or updated very frequently. Primary study selection, where thousands of candidate papers must be reviewed by humans to find the dozens of relevant papers to be included in the SLR, is one of the most time-consuming step in conducting SLRs~\citep{carver2013identifying}.

\subsection{Expert Systems for Primary Study Selection}

The problem of how to efficiently find the dozens of relevant papers among thousands of candidates is categorized as one type of information retrieval problem called {\em total recall}, and has been studied for years~\citep{cormack2015autonomy,cormack2014evaluation,grossman2013,wallace2010semi,miwa2014reducing,Yu2018,Yu2019}. With the goal of optimizing the cost for achieving very high recall--- as close as practicable to 100\%--- with a human assessor in the loop~\citep{grossman2016trec}, the total recall problem can be described as follows~\citep{yu2018total}:

\noindent\begin{minipage}{.99\textwidth}
\begin{RQ}{The Total Recall Problem:}
Given a set of candidates $E$, in which only a small fraction $R \subset E$ are positive, each candidate $x \in E$ can be inspected to reveal its label as positive ($x\in R$) or negative ($x \not\in R$) at a cost. Starting with the labeled set $L = \emptyset$, the task is to inspect and label as few candidates as possible ($\min |L|$) while achieving very high recall ($\max |L\cap R|/|R|$).
\end{RQ}
\end{minipage}

Active learning-based approaches, where machine learning algorithms work alongside humans to learn from human classifications and suggest what needs to be reviewed by humans next, are widely applied in solving total recall problems~\citep{grossman2016trec}. The key idea behind active learning is that a machine learning algorithm can train faster (i.e. using less data) if it is allowed to choose the data from which it learns~\citep{settles2012active}. 
The experience in total recall problems explored to date is that such {\em active learners} outperform supervised and semi-supervised learners and can significantly reduce the effort required to achieve high
\begin{wrapfigure}{r}{1.7in}
\includegraphics[width=1.75in]{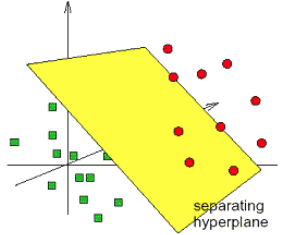}
\caption{Separating positive data points (red circles) from negative ones (green squares).}\label{fig:svm}
\end{wrapfigure}
recall~\citep{Cormack2017Navigating,Cormack2016Engineering,cormack2016scalability,cormack2015autonomy,cormack2014evaluation,grossman2013,wallace2010semi,wallace2010active,wallace2011should,wallace2012class,wallace2013active,Yu2018,Yu2019}. To understand active learning, consider the decision plane
between the positive and negative data points shown in Figure~\ref{fig:svm}. Suppose we want to find more positive data points and we had access to the model shown in Figure~\ref{fig:svm}. One tactic would be to inspect the unlabeled data points that fall into the region of red circles in this figure, as far as possible from the green squares (this tactic is called {\em certainty sampling}). Another tactic would be to verify the position of the boundary, i.e., to inspect the unlabeled data points that are closest to the boundary (this tactic is called {\em uncertainty sampling}). Besides the query tactics, state-of-the-art approaches~\citep{Yu2018,Yu2019} follow the general framework shown in Figure~\ref{fig:{\IT}} and consider the problem of how to stop the inspection at a target recall and how to efficiently correct human errors. When simulated with reverse-engineered primary study selection datasets, these active learning based algorithms can retrieve 90-95\% of relevant papers by reviewing only 5-20\% of the search results~\citep{Yu2019}. This could save weeks of work for humans who might otherwise need to screen thousands of papers.

Despite the foregoing fact, many years have passed while few expert systems have actually been applied in real systematic literature reviews. To the best of our knowledge, only \citet{xiong2018machine} have employed a machine learning aided primary study selection in a systematic review. Even though the machine learning algorithm applied involved a combination of supervised and unsupervised learning (not active learning), they succeeded reducing the cost of primary study selection by around $85\%$. This motivates the study in this article: we would like to conduct a systematic literature review by utilizing a state-of-the-art active learning approach ({\IT}~\citep{Yu2018,Yu2019}) to perform primary study selection. We chose to apply {\IT} to assist the SLR in our case study because in previous work, (1) it outperformed other approaches in terms of inclusion rate~\citep{Yu2018}, (2) its recall estimation provides a confidence for the user that what is missed when the selection stops~\citep{Yu2019}, and (3) it has also been shown effective in solving other software engineering problems~\citep{yu2018total} such as inspecting software security vulnerabilities~\citep{8883076} and finding defect commits~\citep{DBLP:journals/corr/abs-1905-01719}. 

\begin{figure}[!th]
 \centering
 \includegraphics[width=0.9\linewidth]{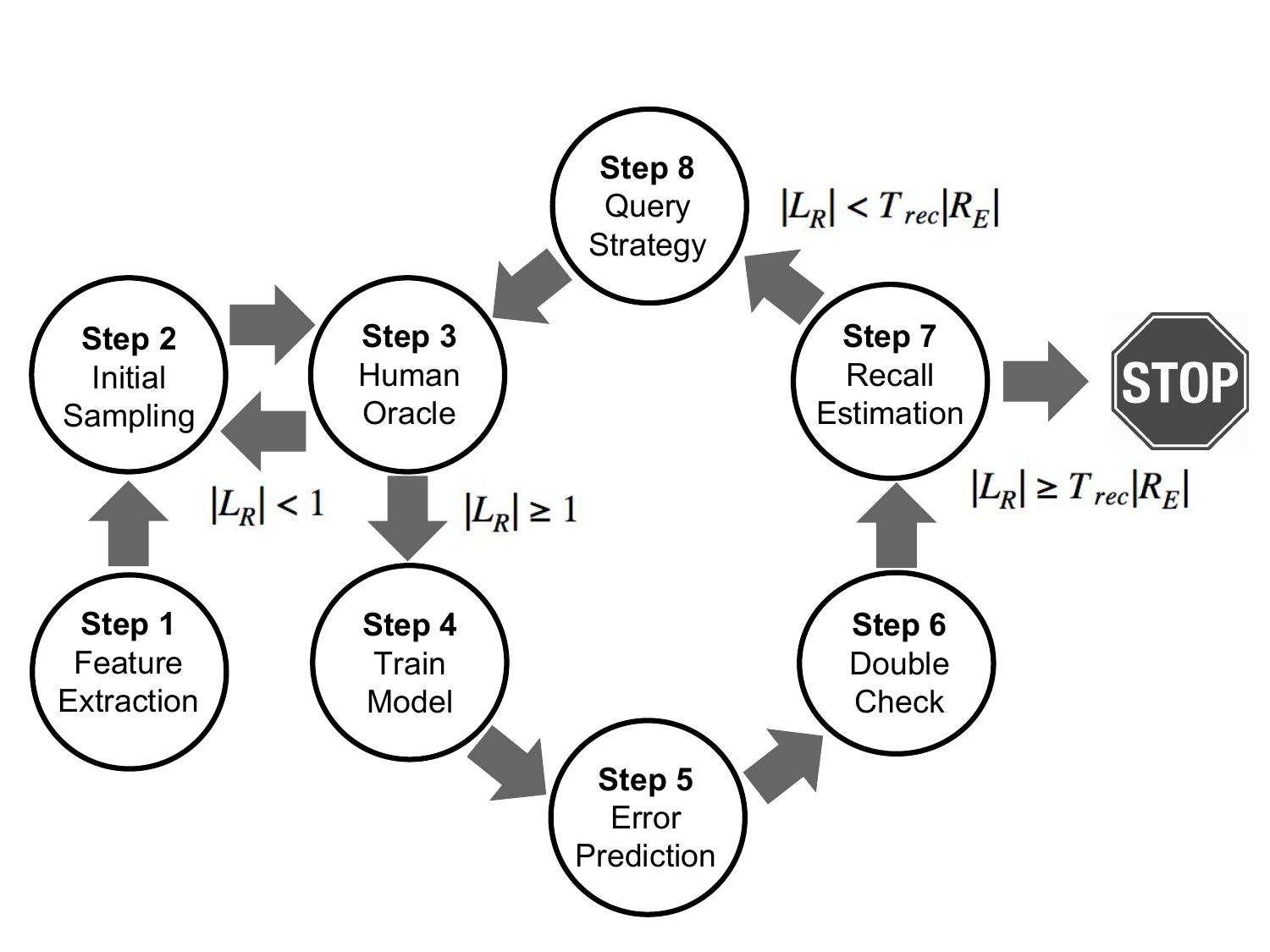}
 \caption{Active learning framework for total recall problems.}
 \label{fig:{\IT}}
\end{figure}

\subsection{{\IT}}

{\IT} is an active learning-based tool\footnote{\url{https://github.com/fastread/src}} that helps reduce the cost of primary study selection in SLRs~\citep{Yu2018,Yu2019}. Consider a primary study selection with
\bi
\item 
$E$: the set of all candidate papers from the search results.
\item
$R$: the set of relevant papers to be included ($R\subset E$).
\item
$L$: the set of papers already reviewed and classified by humans ($L\subset E$).
\item
$L_R = L\cap R$: the set of included papers.
\ei
Instead of reviewing and classifying all candidate papers in a random order, a primary study selection with {\IT} follows the procedure shown in Figure~\ref{fig:{\IT}}, and benefits from three features:
\begin{enumerate}
\item
\textbf{Higher inclusion rate:} {\IT} incrementally trains/updates a machine learning model (in Step 4) on the human classification results ($L$ and $L_R$ from Step 3). With the help of the machine learning model, {\IT} dynamically adjusts the order of papers to be reviewed and classified by humans next (in Step 8) so that relevant papers will be reviewed and included by humans earlier.
\item
\textbf{Recall estimation:} {\IT} estimates the total number of relevant papers $|R_E|\approx |R|$\footnote{Here, $|R_E|\approx |R|$ means that $R_E$ is an estimation of the value of $|R|$.} with a semi-supervised learning algorithm (in Step 7). The human can then stop the primary study selection process when a pre-determined target recall $T_{rec}$ has been reached by estimation $T_{rec}<|L_R|/|R_E|$.
\item
\textbf{Human error correction:} {\IT} also predicts which papers are most likely to have been misclassified by humans (in Step 5). Humans can double check those papers (in Step 6) to correct those errors efficiently.
\end{enumerate}
The pseudo code of {\IT} that we implemented for our case study is shown in Algorithm~\ref{alg:alg} in the Appendix.

\section{Case Study: A Systematic Literature Review on Test Case Prioritization with {\IT}}
\label{sect: Research method}

Changes in a version of a software system may affect the behavior of that system. Regression testing is performed to ensure that changes do not adversely affect the behavior~\citep{catal2013test}. As a regression test suite grows with the size of a software system, software developers need to wait increasingly longer times before they can get useful feedback on their latest commits. In practice, these times can be quite long. As an example, \citet{doi:10.1002/stvr.263} report on a test suite of software with 20,000 lines of code that requires 7 weeks to run. 

Software engineering researchers have explored various techniques for improving the cost-effectiveness of regression testing. Test case prioritization (TCP) is one such technique--- it schedules test cases for execution in an order that attempts to increase their effectiveness at meeting some performance goal~\citep{167}. Unlike other techniques such as test case selection, TCP techniques use the entire test suite and reduce testing cost by achieving parallelization of testing and debugging activities~\citep{22}. By retaining all test cases, TCP techniques do not run the risk of omitting some important test cases.

\begin{figure*}
 \vspace{5mm}
 \centering
 \includegraphics[width=\linewidth]{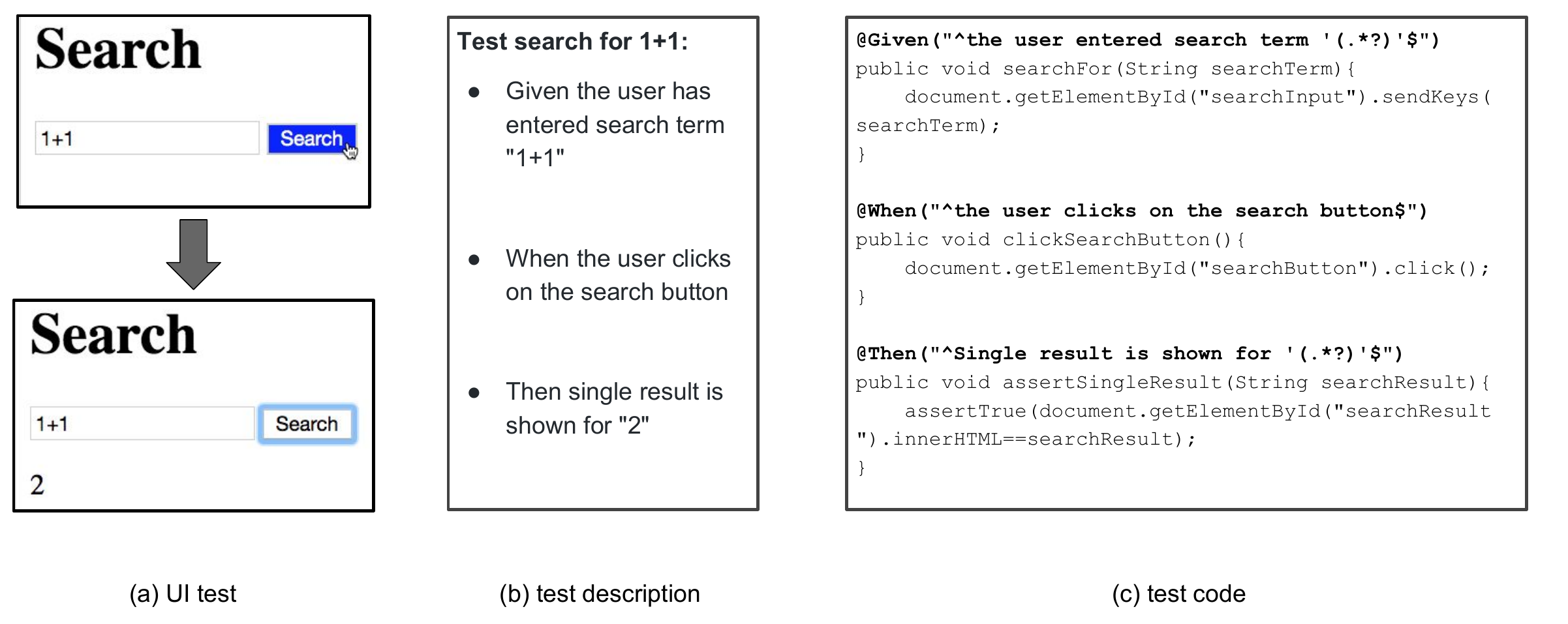}
 \caption{To test a page shown at the left~(a), programmers write a test description~(b) which is converted to test code~(c).}
 \label{fig:example}
\end{figure*}

Automated user interface (UI) testing leads to one special case of regression testing. Compared to unit tests, automated UI tests are more expensive to write and
maintain. Worse still, since automated UI tests are expressed in terms of actions taken by a browser user agent, failures do not have a straightforward relationship to the underlying application code or architecture. Figure~\ref{fig:example} shows how one automated UI test case is designed to exercise a UI performing a simple search on the string ``1+1''. In this example, the test designer wishes to test the search function by (a) verifying that when a user inputs ``1+1'' and clicks the search button, a result of ``2'' will appear. To automate this UI test, the test designer would first (c) define the test code for a set of scenarios, then (b) write the automated UI test case with the pre-defined scenarios and expected input and output. 
In this way, the test designer does not need to know what code will be executed when an automated UI test is executed, and the pre-defined scenarios can be reused in designing other automated UI test cases. As a result, when prioritizing for these automated UI tests, source code information is not available as well as the mapping between a failure of the test case and a fault in the codebase.

We conducted this SLR to identify research papers on TCP techniques that can be applied to automated UI testing. The requirement for such papers is that they can only utilize information available to the prioritization of automated UI tests, i.e. 
\begin{wrapfigure}{r}{2.5in}
\includegraphics[width=2.5in]{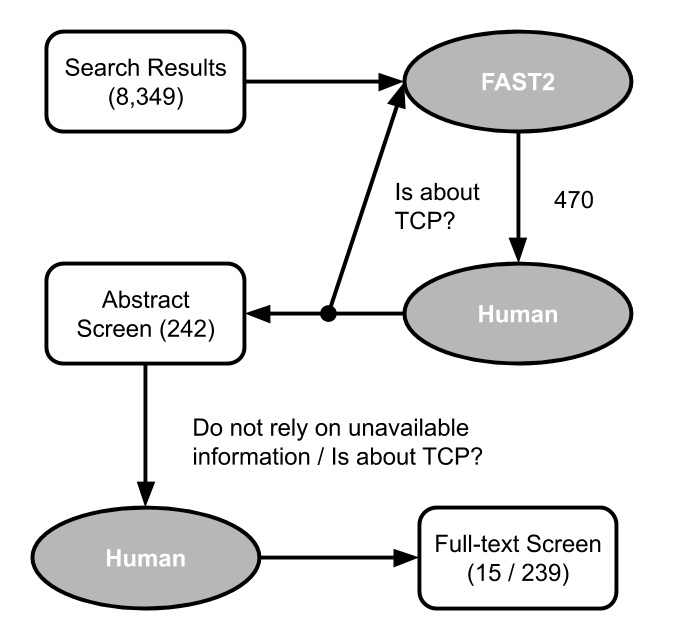}
\caption{Overview of the SLR for finding research papers on TCP techniques that can be applied to automated UI testing.}\label{fig:SLR}
\end{wrapfigure}
(1) description of test cases, (2) historical testing results, and (3) testing results of executed test cases in the current build. The SLR investigated papers from January 1, 1956 to January 1, 2019 and included the following phases, which we go on to describe in turn:
\bi
\item
Planning,
\item
Execution, and
\item
Reporting.
\ei
An overview of the SLR process is shown in Figure~\ref{fig:SLR} where from 8,349 search results, {\IT} cumulatively selected 470 candidate papers for the human to screen. Out of that 470 papers, 242 were included as relevant TCP papers based on titles and abstracts. Then based on full-text screen, 3 were found to be not relevant TCP papers and only 15 were identified as TCP papers that can be applied to automated UI testing, among which 12 TCP techniques were summarized.

\subsection{Phase 1--- planning}
\label{sect: planning}

In this phase, we specified research questions, search strategy, inclusion and exclusion criteria, classification of papers, and threats to validity.

\subsubsection{Research questions}

The only research question of this case study is:
\begin{highlight*}
\textbf{What test case prioritization algorithms utilize only (1) description of test cases, (2) historical testing results, and (3) testing results of executed test cases in the current build information?}
\end{highlight*}

\subsubsection{Search strategy}
\label{sect: search strategy}

In the search process, we want to first find all primary studies on test case prioritization. With Boolean operator OR to link the synonyms of the main terms and Boolean operator AND to combine the main terms, the search string we applied is as follows:

\emph{[software AND test AND (rank* OR optimi* OR prioriti*)]}.

\noindent We executed this search string in the IEEE Explore\footnote{https://ieeexplore.ieee.org} database to find papers containing the keywords in their titles and abstracts. We chose to search IEEE Explore because it covers a large portion of the software engineering publications and is the only database we know of in which thousands of search results can be downloaded automatically with their titles and abstracts. {The ``automated UI testing'' keywords were not used in this search since (1) it literally generated no result on IEEE Xplore (too narrow), and (2) most test case prioritization techniques should be able to apply to automated UI testing if they are not using the source code related information. }

\subsubsection{Inclusion and exclusion criteria}

This review included papers on test case prioritization published between 1956 and 2018. Papers from peer-reviewed
journals, conferences, and workshops were considered. We excluded papers that were not related to test case prioritization in the context of software engineering, such as papers on test case selection or fault localization. The inclusion criteria (IC) and exclusion criteria (EC) are as follows:

\begin{enumerate}[leftmargin=26pt,start=1,label={\bfseries IC \arabic*}]
\item
Primary papers on TCP.
\item
Secondary papers on TCP.
\end{enumerate}
\begin{enumerate}[leftmargin=26pt,start=1,label={\bfseries EC \arabic*}]
\item
Primary papers on test case selection or test suite reduction only.
\item
Primary papers on test case generation only.
\item
Primary papers on fault localization.
\end{enumerate}
Primary study selection was performed by the first author alone. {\IT} was applied to help this process include 90\% of the relevant papers ($T_{rec} = 0.9$, $N_1 = 10$, $N_2 = 30$\;). We targeted 90\% recall because the creators of {\IT} suggest that 90-95\% recall is appropriate because the cost required to reach higher recall increases exponentially~\citep{Yu2018,Yu2019}. Whether 90\% recall is in fact sufficient will be examined further in Section~\ref{sect: validate}.

\subsubsection{Classification of papers}
\label{sect: Classification of papers}

Classification was also performed by the first author alone. The papers were classified according to what information they utilized during the prioritization. Details on each category will be provided in Section~\ref{sect: reporting}. These categories for are non-exclusive. For example, one paper may utilize both source code and history information.

\subsubsection{Threats to validity}
\label{sect: Threats to validity}

There are two major validity threats to this systematic literature review:
\begin{enumerate}
\item
Only one data source: we searched for papers in one data source (IEEE Xplore) because retrieving search results in other databases would have been inordinately expensive. Therefore, TCP papers in journals or conference proceedings that are not indexed by IEEE Xplore were not included in this SLR study. 
\item
Primary study selection with {\IT}: this is the first SLR study conducted with {\IT}, applied to a single case, and the extent to which results will generalize cannot be determined.
\end{enumerate}

\subsection{Phase 2--- execution}
\label{sect: execution}

\subsubsection{Search}

After applying the search string discussed in Section~\ref{sect: search strategy} in IEEE Xplore, we obtained a result of 8,381 candidate papers. These 8,381 papers were downloaded automatically with their title, abstract, pdf link, and publication year information. Among the 8,381 papers, 32 were not research papers (e.g., they were editorials or prefaces) and were thus excluded. The search process, including the design of the search string and retrieval of all the search result, required approximately 1 hour.


\subsubsection{Primary study selection}

\begin{figure}[!tb]
 \centering
 \includegraphics[width=\linewidth]{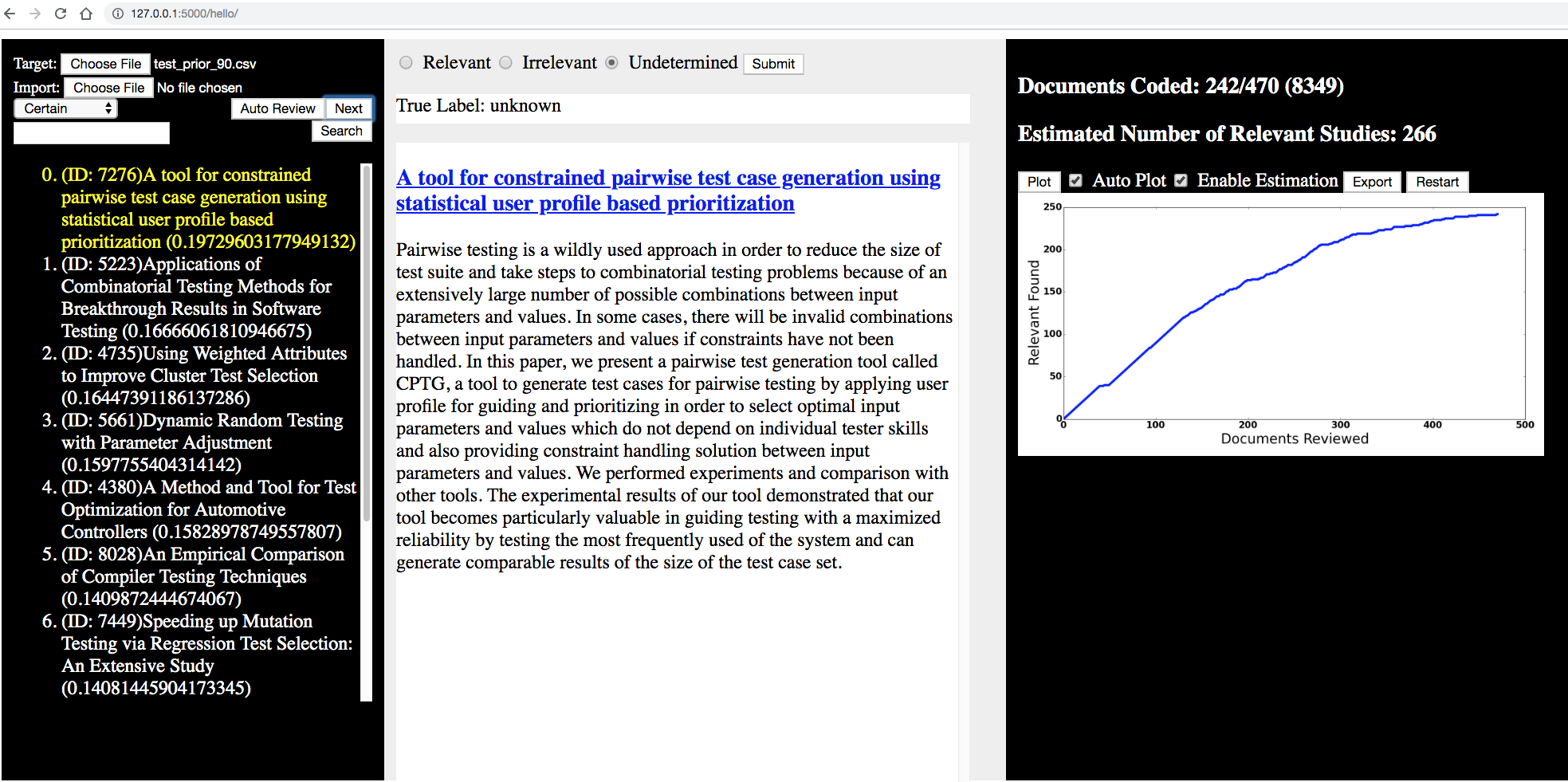}
 \caption{Interface of the {\IT} tool when the study selection stopped.}
 \label{fig: tool}
\end{figure}

Following the instructions for using the {\IT} tool\footnote{\url{https://github.com/fastread/src}}, the first author performed the following steps to select the primary studies with a target recall $T_{rec}=90\%$:
\begin{enumerate}
\item
The first author loaded the search results of 8,349 papers with their titles and abstracts into {\IT}.
\item
The first author \textbf{searched} for keywords ``test prioritization'' and screened the first ten results by reading the titles and abstracts. Ten papers were included as relevant.
\item
Given that $|L_R|=10\ge 1$, when the \textbf{Next} button is selected, an SVM model is trained based on the ten screened papers and suggestions for uncertainty sampling and certainty sampling are provided. Because $|L_R|=10< 30$, the papers suggested during uncertainty sampling were screened.
\item
After 20 more papers were reviewed based on uncertainty sampling (the SVM model was retrained and suggestions were updated for every 10 papers screened by the author), 30 relevant and zero non-relevant papers were screened. Because $|L_R|=30\ge 30$, certainty sampling was applied for the rest of the papers.
\item
Finally, when 440 more papers had been screened based on certainty sampling, 242 relevant papers were found among the 470 reviewed ones ($|L_R|=242$ and $|L|=470$), as shown in Figure~\ref{fig: tool}. Meanwhile, the estimated recall was $|L_R|/|R_E|=242/266=91\%$. This was the first time the estimated recall reached the target recall ($T_{rec}=90\%$). The selection thus stopped and the results were exported.
\end{enumerate}
The primary study selection process with {\IT} required approximately three hours of effort by one person.

\subsubsection{Full-text review and paper classification}

The 242 papers identified in the search were reviewed in full text and classified according to the information utilized. Among the 242 papers, three were determined to be not relevant (not about test case prioritization) based on their full text and were thus excluded. Then, the rest 239 papers were classified regarding to the information utilized and only 15 papers were found to be applicable to automated UI testing. The paper classification process required approximately 40 hours of effort by one person. Details of the classification process will be introduced in the next subsection.

\subsection{Phase 3--- reporting}
\label{sect: reporting}

\begin{figure}[!t]
 \centering
 \includegraphics[width=.8\linewidth]{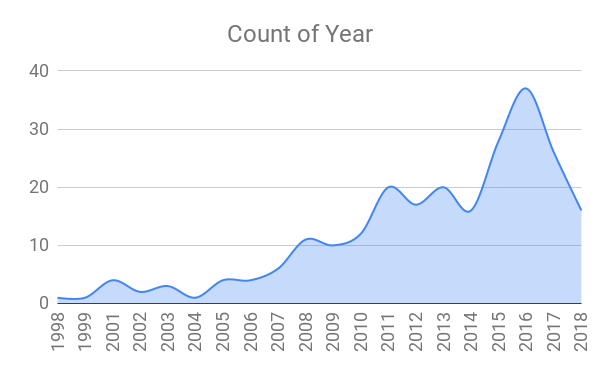}
 \caption{Distribution of publication years}
 \label{fig:year}
\end{figure}

In this phase, the analytical results of the systematic literature review are discussed. These results are collected by a full-text reviewing of the 239 papers identified in the search. The distribution of publication years on the selected papers is shown in Figure~\ref{fig:year}.

\begin{figure}[!t]
 \centering
 \includegraphics[width=.8\linewidth]{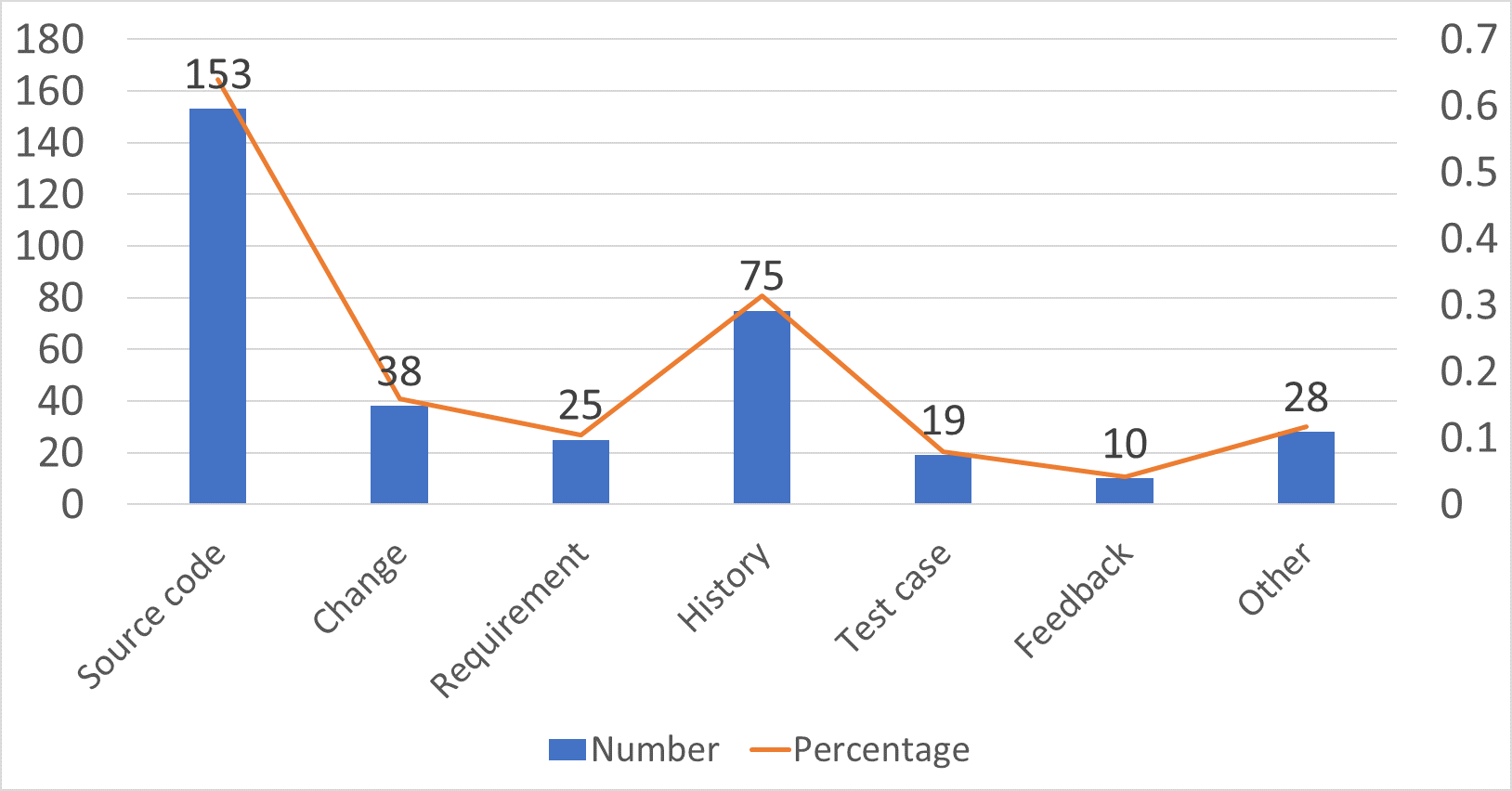}
 \caption{Distribution of information used}
 \label{fig:information}
\end{figure}

To find out which TCP algorithms can be applied to prioritize automated UI tests, we first categorize each primary study by the information utilized. Figure~\ref{fig:information} shows the distribution of each category of information being used by the techniques presented in the 239 papers. The seven categories of information we analyzed are listed in the table as follows:
\bi
\item
\textbf{Source code:} source code under test. About half of the analyzed TCP methods extract features from the source code being tested, e.g. software metrics~\citep{129}, code coverage~\citep{34}. 
\item
\textbf{Change:} code change from the prior build. Around 16\% of the analyzed papers present TCP techniques that utilize information about "what has been changed in the source code" to decide which test cases should be executed earlier~\citep{130}.
\item
\textbf{Requirement:} requirements properties. Around 10\% of the analyzed papers present TCP techniques that utilize this type of information, such as customer-assigned priority on requirements, requirement volatility, or developer-perceived implementation complexity of requirements, for prioritization~\citep{57}.
\item
\textbf{History:} execution results (pass/fail/skip) from previous runs. About 30\% of the analyzed papers present TCP techniques that utilize history information such as the fault/failure exposing potential of each test case to reorder the test cases~\citep{84}.
\item
\textbf{Test cases:} information about the test cases, e.g. test descriptions, test code, etc. Some papers (17) present TCP techniques that utilize this type of information to calculate the similarity between test cases, then prioritize the test cases based on the similarity and other types of information~\citep{195}.
\item
\textbf{Feedback:} execution results (pass/fail/skip) of test cases on current run. A few (10) papers present TCP techniques that learn from the results of already executed test cases to dynamically re-prioritize test cases that have not yet been executed~\citep{103}.
\item
\textbf{Other:} other information such as test dependency~\citep{8} or test impact~\citep{203}.
\ei

From Figure~\ref{fig:information} we see that source code is the most frequently used information. What we also found is that most of the papers (around 50\%) present coverage-based methods, which prioritize test cases in orders that reach maximum coverage with minimum testing cost, by using greedy or search-based algorithms~\citep{69}. However, these coverage-based methods are not applicable when information such as source code, change, and requirement is not available. Additionally, although history information is also frequently used, it is actually usually being used along with other information such as source code. As a result, our next step is to identify the TCP algorithms only utilizing history, test case, and feedback information. This gives us only 15 research papers. Among these 15 papers, we identified 12 state-of-the-art TCP techniques that can be applied to automated UI testing, after accounting for similar algorithms and removing inapplicable ones. The 12 TCP techniques thus identified are grouped by the information they use and are listed in Table~\ref{tab:algorithms}:

\bi
\item
\textbf{History-based (B and C):} only history information is utilized. B group algorithms apply different metrics extracted from past test execution results (pass/fail/skip) to predict fault/failure exposing potential, and to prioritize test cases~\citep{38,79,7,184,212,232,97,124,151,182}. C1 is different from B group algorithms since C1 only utilizes the history information to estimate the runtime of each test case~\citep{124,184,213}.
\item
\textbf{Test case-based (D):} test case information and history information is utilized. D group algorithms utilize test case information to calculate the similarity between test cases, and then prioritize the test cases based on both test case similarity and history information~\citep{18,59}.
\item
\textbf{Feedback-based (E):} feedback and history information is utilized. E group algorithms utilize the execution results (pass/fail/skip) in a {\em current} build to dynamically reorder those test cases that have not yet been executed~\citep{103,132,183}.
\ei
In our prior work~\citep{yu2019terminator} we then applied these techniques as baselines and compared them against a proposed new TCP algorithm on datasets used in automated UI testing at LexisNexis. Detailed descriptions of how each algorithm work can also be found in that prior work.

\begin{table*}
\caption{Eligible Test Case Prioritization Techniques and Information They Utilize}
\label{tab:algorithms}
\begin{center}
\footnotesize
\setlength\tabcolsep{4pt}
\begin{tabular}{c|p{6.5cm}|P{1.5cm}P{1.6cm}P{1.6cm}|}
\multicolumn{2}{c|}{~} & \multicolumn{3}{c|}{\textbf{Utilized Information}}\\\cline{3-5}
\textbf{ID} & \textbf{Technique Description} & Execution history & Test case description & Feedback \\ \hline
B1 & Execute test cases in ascending order of time since last failure~\citep{38,79}. & \checkmark & & \\ 
B2 & Execute test cases in descending order of number of times failed/number of times executed~\citep{7,184,212,232}. & \checkmark & & \\ 
B3 & Execute test cases in descending order of exponential decay metrics~\citep{97,124,151}. & \checkmark & & \\ 
B4 & Execute test cases in descending order of ROCKET metrics~\citep{182}. & \checkmark & & \\ 
B5 & Execute test cases in descending order of the Mahalanobis distance of each test case to the origin (0,0) when considering two metrics--- time since last execution and failure rate~\citep{232}. & \checkmark & & \\\hline
C1 & Execute test cases in ascending order of the estimated test case runtime~\citep{124,184}. & \checkmark & & \\ \hline
D1 & Supervised learning with Simple History (SH)~\citep{59}. & \checkmark & \checkmark & \\ 
D2 & Supervised learning with All History (AH)~\citep{59, 18}. & \checkmark & \checkmark & \\ 
D3 & Supervised learning with Weighted History (WH)~\citep{59}. & \checkmark & \checkmark & \\ \hline
E1 & Dynamic test case prioritization with co-failure information~\citep{183}. & \checkmark & & \checkmark \\ 
E2 & Dynamic test case prioritization with flipping history~\citep{103}. & \checkmark & & \checkmark \\ 
E3 & Dynamic test case prioritization with rules mined from failure history~\citep{132}. & \checkmark & & \checkmark \\ \hline
\end{tabular}
\end{center}

\end{table*}




\section{Validation}
\label{sect: validate}

{We now address the research questions, by first validating the primary study selection results of {\IT} with a controlled experiment of the relevant paper selection process. Six graduate students (from our lab with at least 1 year of software engineering research experience) were enlisted to manually screen the candidate papers. A full-text review on the relevant papers included by either the six students or {\IT} were performed to provide the final decision on which papers should be included. }

\subsection{RQ1: What percentage of relevant papers did {\IT} actually retrieve?}
\label{sect: RQ1}

Considering the prohibitive cost involved in manually screening 8,349 papers, we validate our results on only a subset of the candidate papers. By searching in IEEE Xplore with the following search string

\emph{[software AND test AND prioriti*]}.

\noindent a validation set of 783 papers was retrieved. Among the 470 of these papers that had been screened with {\IT}, 318 were in the validation set. Among the 242 papers classified as relevant by {\IT}, 237 were in the validation set.

Each paper in the validation set was manually screened by at least two graduate students. A third student was asked to screen the paper if the screening results from the first two were inconsistent. A majority vote was then used to determine the final screening results of papers (293 relevant papers) in the validation set. After that, full-text validation was applied to the papers identified by the majority vote that were not identified by {\IT}. {Six hours were spent on the full-text validation of the 70 papers, and it was confirmed that 39 of the validated papers were relevant.} These full-text validation results were treated as the ground truth for the validation set.

\begin{table}[!tb]
\caption{Primary Study Selection Validation Results}
\label{tab:validate}
\centering
\setlength\tabcolsep{6pt}
\begin{tabular}{p{1.5cm}|p{0.5cm}|P{1cm}P{1cm}P{1.2cm}|P{1cm}P{1cm}|P{1cm}}
\toprule
\multicolumn{2}{c|}{} & \multicolumn{3}{c|}{{\IT}} & \multicolumn{2}{c|}{Majority Vote} & \\ \cline{3-7}
\multicolumn{2}{c|}{} & yes & no & ignored & yes & no & \\\hline
Ground & yes & 234 & 12 & 27 & 259 & 14 & 273\\
Truth & no & 3 & 69 & 438 & 34 & 476 & 510\\\hline
\multicolumn{2}{c|}{} & 237 & 81& 465& 293 & 490 & 783 \\\bottomrule
\end{tabular}
\end{table}

\begin{table}[!tb]
\caption{Labels of Table~\ref{tab:validate}}
\label{tab:labels}
\centering
\begin{tabular}{p{2.2cm}|P{1.3cm}|p{9.5cm}}
\toprule
\multicolumn{2}{c|}{} & \makecell{\centering Description} \\\hline
 & yes & Papers suggested by {\IT} and included by human \\
{\IT} & no & Papers suggested by {\IT} but excluded by human\\
 & ignored & Papers ignored by {\IT}\\\hline
Majority & yes & Papers included by two humans\\
Vote & no & Papers excluded by two humans\\\hline
Ground & yes & Papers included by full-text validation\\
Truth & no & Papers excluded by full-text validation or by both {\IT} and majority vote\\\bottomrule
\end{tabular}
\end{table}

Table~\ref{tab:validate} summarizes the validation results of {\IT}, the majority vote, and the ground truth with labels explained in Table~\ref{tab:labels}. From Table~\ref{tab:validate} we can derive the performance of {\IT} on the validation set:
\bi
\item
$\text{Human Precision} = \frac{234}{234+3} = 0.99 $
\item
$\text{{\IT} Precision} = \frac{234}{237+81} = 0.74 $
\item
$\text{Recall} = \text{{\IT} Recall} \times \text{Human Recall} = \frac{234+12}{273}\times \frac{234}{234+12} = 0.90\times0.95 = 0.85$
\item
$\text{Cost} = \frac{237+81}{783} = 0.41$
\ei
Here, the recall involved in selecting primary studies consists of two parts--- {\IT} recall and human recall. The {\IT} recall on the validation set is 90\%, which is very close to its estimation of $91\%$ recall and is the same as the target recall $T_{rec}=90\%$. Therefore, we conclude that the recall estimation of {\IT} was accurate in this SLR study.

As for the performance of manual screening with majority votes, the following calculations apply:
\bi
\item
$\text{Precision} = \frac{259}{259+34} = 0.88 $
\item
$\text{Recall} = \frac{259}{259+14} = 0.95$
\item
$\text{Cost} = \frac{783\times 2+174}{783} = 2.22$
\ei
This data shows that the human working with {\IT} (the first author) achieved the same recall as, but higher precision ($99\%$) than, the majority vote results of the other six humans ($89\%$). This probably occurred because the author designed the inclusion and exclusion criteria and had a better understanding of which papers are relevant to the SLR. This result suggests that, although employing more human reviewers for relevant paper selection can effectively reduce the time required for that process, more cost-effective and precise results can be achieved if only the human planning the SLR is employed for the primary study selection, which leads to less unnecessary full-text review effort.

\begin{figure}[!t]
 \centering
    \subfloat[Number]
    {
        \includegraphics[width=.8\linewidth]{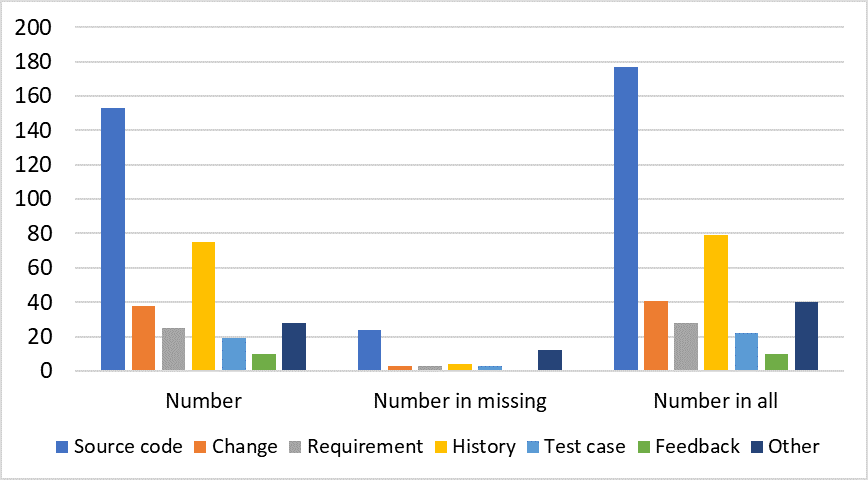}
    }\\
    \subfloat[Percentage]
    {
        \includegraphics[width=.8\linewidth]{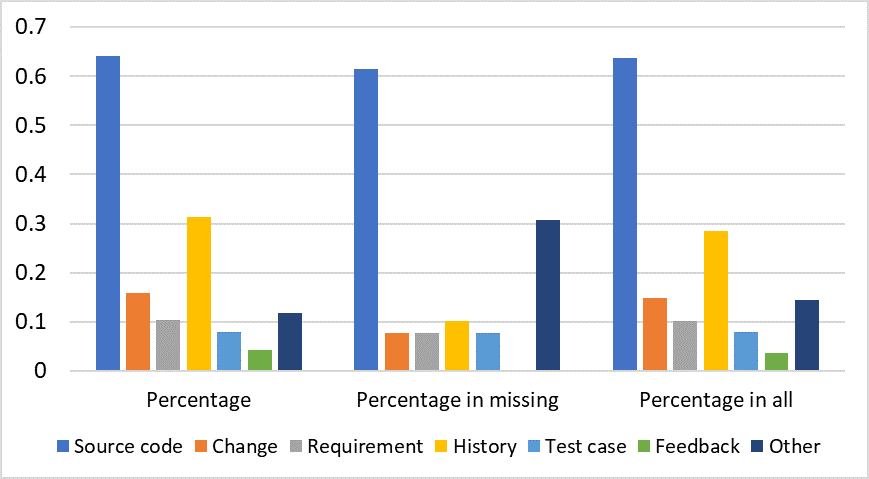}
    }
 \caption{Comparing the distribution of papers in each category (of information utilized), considering (1) the 239 papers considere relevant by {\IT} (Percentage), (2) the 39 relevant papers missed by {\IT} (Percentage in missing), and (3) the 278 ground truth relevant papers (Percentage in all).}
 \label{fig:missing}
\end{figure}

\subsection{RQ2: What information is missing in the final report because of {\IT}?}
\label{sect: RQ2}

To determine what information is lost by excluding the 39 relevant papers not discovered by {\IT} and the human reviewer, we analyzed these papers in the same manner as the 239 papers that were initially included, and classified them based on prioritization goals, data types, information used, and method applied for black box testing. As shown in Figure~\ref{fig:missing}, we observe:
\bi
\item
Distributions of the missing papers into categories are quite different from those obtained for the 239 papers identified as relevant by {\IT}. This suggests that when using {\IT}, a bias could be introduced in terms of which relevant papers will be retrieved. This is probably caused by the imbalance of categories in the training data of {\IT}, e.g., when $30\%$ of the training data (relevant papers found) uses history information, it is likely that {\IT} would predict that a paper using history information has a higher probability of being relevant than otherwise.
\item
Distributions of all the relevant papers into categories are still similar to those obtained for the 239 papers identified as relevant by {\IT} -- especially with respect to the rankings of number of papers in each category. This suggests that, despite the bias introduced by {\IT}, the overall conclusions of the SLR are still representative when using the $90\%$ relevant papers selected with {\IT}. However, in studies like systematic mappings where the exact values of distributions matter, such biases should be avoided by manually screening all of the search results.
\item
While one more paper~\citep{552} in the papers omitted by {\IT} was identified to be applicable to automated UI testing, that technique is similar to the D2 technique~\citep{59} listed in Table~\ref{tab:algorithms}. Therefore, including that paper did not add any new information to the conclusions of the SLR case study.
\ei

To summarize, the benefits associated with the use of {\IT} to guide the selection of relevant papers are as follows:
\bi
\item
With the help of {\IT}, 85\% of the relevant papers were included with only $\frac{470}{8349}=6\%$ of the candidate papers screened. This saved approximately 50 hours of work. 
\item
The recall of {\IT} when the selection stopped was close to the target recall based on the validation result. This suggests that a researcher may be able to choose a level of recall at which to stop the selection with the help of the recall estimation given by {\IT}.
\item
The cost of primary study selection was reduced to a reasonable level (three hours for 470 papers), so that it was possible to employ only one human (the one who planned the SLR) to select relevant papers. This reduced the human error rate for the selection process.
\ei
The costs associated with use of {\IT} to guide the selection of relevant papers are as follows:
\bi
\item
There were more missed relevant papers when applying {\IT} with a target recall lower than 100\%. The higher the target recall is, the higher the cost will be~\citep{Yu2019}. Researchers need to consider the tradeoff between the screening effort they are willing to spend and the recall they can achieve with {\IT}.
\item
Using {\IT} can introduce a sampling bias into the included relevant papers. This may not always affect SLR studies (e.g. the conclusions of the case study SLR in this paper remained unchanged) but it should be avoided in studies such as systematic mappings.
\ei

\section{Conclusions and Future Work}
\label{sect: Conclusions and Future Work}

When researchers report work involving the creation of new approaches, they usually focus on the designed novel approach and how that approach performed compares to other state-of-the-art approaches. However, they seldom discuss how they find these state-of-the-art baseline approaches. In this paper, we want to bring researchers' attention to this problem--- how to find the state-of-the-art baseline approaches--- because (1) it is critical to any research (researchers can keep improving the solution to a problem only when they are fully aware of what others have done to attempt to solve that problem), and (2) it is usually a tedious task that consumes large amounts of time and effort. Expert systems have been designed to assist researchers more efficiently in finding relevant papers. However, these expert systems have not been widely applied because of the lack of successful literature review case studies supporting their use. 

To this end, we report the process of finding 12 state-of-the-art baseline test case prioritization algorithms, in our lately published test case prioritization paper~\citep{yu2019terminator}. In this literature review case study, we investigated 242 papers on test case prioritization that had been published in conference proceedings and journals, through a systematic literature review process. The 242 papers were selected by manually screening 470 of the 8,349 candidate papers with the help of {\IT}, a human-in-the-loop expert system. During this process, {\IT} reduced the effort required for paper selection by $1-\frac{470}{8349}=94\%$ (from 53 hours to three hours). Based on the validation results in which six other humans screened a subset (783) of the candidate papers, the {\IT} selection process included $85\%$ of the relevant papers with human errors contributing to $5\%$ of the missing papers. Given the large reduction of human effort required for primary study selection, with only $10\%$ loss in recall (the same as the {\IT}'s estimation), and given the fact that the missing relevant papers did not affect the final conclusions of the case study SLR, this work supports the suggestion that {\IT} can be used to cost-effectively select primary studies in SLRs. We did find, however, that using {\IT} can introduce a sampling bias in the included relevant papers. Thus, when conducting systematic mapping studies, it may be best to avoid using {\IT}.

For future work, we intend to encourage other software engineering researchers to conduct systematic literature reviews using {\IT}. We also intend to find ways to improve the efficiency (higher recall and lower cost) of our expert system-assisted primary study selection approach through simulations on the SLR datasets including this study.
Finally, we will attempt to alleviate the sampling bias introduced by {\IT}. A possible solution in this context may be to replace {\IT}'s learner with some instance-based classifiers such as K-Nearest Neighbors.

\section*{Acknowledgement}

We thank Tianpei Xia, Huy Tu, Jianfeng Chen, Xueqi Yang, Rui Shu and Fahmid Morshed Fahid for their effort in performing the controlled experiment of relevant paper selection in Section~\ref{sect: validate}.
 
\vspace*{0.5mm}

\balance

\newpage

\bibliography{sigproc,literature} 

\newpage
\appendix
\section{Pseudo Code}
\label{sect:appendix}

\begin{algorithm}[!tbh]
\scriptsize
\SetKwInOut{Input}{Input}
\SetKwInOut{Output}{Output}
\SetKwInOut{Parameter}{Parameter}
\SetKwRepeat{Do}{do}{while}
\Input{$E$, the candidate paper set (search results)\\$T_{rec}$, target recall\\$N_1$, batch size\\$N_2$, threshold of query strategy\\$Q$, search query for BM25 to boost initial selection}
\Output{$L$, screened papers\\$L_R$, included relevant papers}
\BlankLine
$L\leftarrow \emptyset$\; $L_R\leftarrow \emptyset$\; $|R_E|\leftarrow \infty$\;
\BlankLine
\tcp{Keep screening until target recall $T_{rec}$ has been achieved.}
\While{$|L_R| < T_{rec}|R_E|$}{
 \tcp{Start training when first relevant paper is found}
 \eIf{$|L_R| \geq 1$}{
 \tcp{Alleviate bias in negative training examples}
 $L_{\text{pre}}\leftarrow \text{Presume}(L,E\setminus L)$\;
 $CL\leftarrow \text{Train}(L_{\text{pre}})$\;
 \tcp{Estimate \#relevant papers}
 $|R_E|\leftarrow SEMI(CL,E,L,L_R)$\;
 \tcp{Select unscreened papers for human to screen}
 $X\leftarrow \text{Query}(CL,E\setminus L,L_R)$\;
 }{
 \tcp{Select unscreened papers for human to screen by BM25 Ranking}
 $X\leftarrow argsort(BM25(E\setminus L,Q))[:N_1]$\;
 }
 \tcp{Human screen selected papers}
 \ForEach{$x \in X$}{
 \tcp{Include paper if relevant}
 \If{\text{Screen}(x)}{
 $L_R\leftarrow L_R \cup x$\;
 }
 \tcp{Add paper into screened set}
 $L\leftarrow L \cup x$\;
 }
}
\Return{$L,L_R$}\;
\BlankLine
\Fn{Presume ($L,E\setminus L$)}{
 \tcp{Randomly sample $|L|$ points from $E\setminus L$, presume those as non-relevant}
 \Return $L \cup \text{Random}(E\setminus L,|L|)$\;
}
\BlankLine
\Fn{Train ($L_{\text{pre}}$)}{
 \tcp{Train linear SVM with Weighting}
 $CL\leftarrow \text{SVM}(L_{\text{pre}},\text{kernel=linear},\text{class}\_\text{weight=balanced})$\;
 \If{$L_R\ge N_2$}{
 \tcp{Aggressive undersampling}
 $L_I\leftarrow L_{\text{pre}}\setminus L_R$\;
 $\text{tmp}\leftarrow L_I[\text{argsort}(CL.\text{decision}\_\text{function}(L_I))[:|L_R|]]$\;
 $CL\leftarrow \text{SVM}(L_R \cup \text{tmp, kernel=linear})$\;
 }
 \Return{$CL$}\;
}
\caption{Pseudo Code for {\IT}~\citep{Yu2019} Implemented in the SLR}\label{alg:alg}
\end{algorithm}
\begin{algorithm}[!tbh]
\scriptsize
\BlankLine
\Fn{Query ($CL,E\setminus L,L_R$)}{
 \eIf{$L_R\ge N_2$}{
 \tcp{Certainty Sampling (highest predicted probability of failing)}
 $X\leftarrow \text{argsort}(CL.\text{decision}\_\text{function}(E\setminus L))[::-1][:N_1]$\;
 }{
 \tcp{Uncertainty Sampling}
 $X\leftarrow \text{argsort}(\text{abs}(CL.\text{decision}\_\text{function}(E\setminus L)))[:N_1]$\;
 }
 \Return{$X$}\;
}
\BlankLine
\Fn{Screen ($x$)}{
 \eIf{\text{human thinks x is relevant}}{
 \Return{\text{True}}\;
 }{
 \Return{\text{False}}\;
 }
}
\BlankLine
\Fn{SEMI ($CL,E,L,L_R$)}{
 $|R_E|_{last}\leftarrow 0$\;
 $\neg L \leftarrow E \setminus L$\;
 \ForEach{$x \in E$}{
 $D(x) \leftarrow CL.decision\_function(x)$\;
 \If{$x \in |L_R|$}{
 $Y(x)\leftarrow 1 $\;
 }
 \Else{
 $Y(x)\leftarrow 0 $\;
 }
 }
 $|R_E| \leftarrow \sum\limits_{x\in E} Y(x)$\;
 \While{$|R_E|\neq |R_E|_{last}$}{
 \tcp{Fit and transform Logistic Regression}
 $LReg \leftarrow LogisticRegression(D,Y)$\;
 $Y \leftarrow TemporaryLabel(LReg,\neg L,Y)$\;
 $|R_E|_{last}\leftarrow |R_E|$\;
 \tcp{Estimation based on temporary labels}
 $|R_E| \leftarrow \sum\limits_{x\in E} Y(x)$\;
 }
 \Return{$|R_E|$}\;
}
\BlankLine
\Fn{TemporaryLabel ($LReg,\neg L,Y$)}{
 $count \leftarrow 0$\;
 $target \leftarrow 1$\;
 $can \leftarrow \emptyset$\;
 \tcp{Sort $\neg L$ by descending order of $LReg(x)$}
 $\neg L \leftarrow SortBy(\neg L,LReg)$\;
 \ForEach{$x \in \neg L$}{
 $count \leftarrow count+LReg(x)$\;
 $can \leftarrow can \cup \{x\}$\;
 \If{$count \geq target$}{
 $Y(can[0]) \leftarrow 1$\;
 $target \leftarrow target+1$\;
 $can \leftarrow \emptyset$\;
 }
 }
 \Return{$Y$}\;
}
\end{algorithm}

\end{document}